\documentclass[5p]{elsarticle}

\usepackage{amssymb}

\journal{TIPP09 Proceedings in NIMA}

\begin{document}

\begin{frontmatter}



\title{Cosmic Ray Tests of the Prototype TPC for the ILC Experiment} 


%
%
%
%
%
\author[4]{K.~Ackermann}
\author[10]{S.~Arai}
\author[7]{D.C.~Arogancia}
\author[7]{A.M.~Bacala}
\author[1]{M.~Ball}
\author[1]{T.~Behnke}
\author[10]{H.~Bito}
\author[4]{V.~Eckardt}
\author[13]{K.~Fujii}
\author[14]{T.~Fusayasu}
\author[1]{N.~Ghodbane}
\author[7]{H.C.~Gooc Jr.}
\author[10]{T.~Kijima}
\author[1]{M.~Hamann}
\author[10]{M.~Habu}
\author[3]{R.-D.~Heuer}
\author[8]{K.~Hiramatsu}
\author[13]{K.~Ikematsu}
\author[5]{A.~Kaukher}
\author[9]{H.~Kuroiwa}
\author[1]{M.E.~Janssen}
\author[8]{Y.~Kato}
\author[13]{M.~Kobayashi\corref{cor1}}
     \ead{makoto.kobayashi.exp@kek.jp}
     \cortext[cor1]{Corresponding author.
                           Tel.: +81 29 864 5379; fax: +81 29 864 2580.}
\author[1]{T.~Kuhl}
\author[3]{T.~Lux}
\author[13]{T.~Matsuda}
\author[10]{S.~Matsushita}
\author[10]{A.~Miyazaki}
\author[10]{K.~Nakamura}
\author[10]{O.~Nitoh}
\author[10]{H.~Ohta}
\author[7]{R.L.~Reserva}
\author[10]{K.~Sakai}
\author[9]{N.~Sakamoto}
\author[11]{T.~Sanuki}
\author[4]{R.~Settles}
\author[9]{A.~Sugiyama}
\author[6]{T.~Takahashi}
\author[10]{T.~Tomioka}
\author[9]{H.~Tsuji}
\author[12]{T.~Watanabe}
\author[1]{P.~Wienemann}
\author[1]{R.~Wurth}
\author[9]{H.~Yamaguchi}
\author[9]{M.~Yamaguchi}
\author[15]{A.~Yamaguchi}
\author[11]{T.~Yamamura}
\author[13]{H.~Yamaoka}
\author[8]{T.~Yazu}
\author[2]{R.~Yonamine}
%
%
\address[4]{Max-Planck-Institute for Physics, Munich, Germany}
\address[10]{Tokyo University of Agriculture and Technology, Tokyo, Japan}
\address[7]{Mindanao State University, Iligan City, Philippines}
\address[1]{DESY, Hamburg, Germany}
\address[13]{KEK, IPNS, Ibaraki, Japan}
\address[14]{Nagasaki Institute of Applied Science, Nagasaki, Japan}
\address[3]{Universit\"{a}t Hamburg, Hamburg, Germany}
\address[8]{Kinki University, Osaka, Japan}
\address[5]{Universit\"{a}t Rostock, Rostock, Germany}
\address[9]{Saga University, Saga, Japan}
\address[11]{University of Tokyo, ICEPP, Tokyo, Japan}
\address[6]{Hiroshima University, Hiroshima, Japan}
\address[12]{Kogakuin University, Tokyo, Japan}
\address[15]{University of Tsukuba, Ibaraki, Japan}
\address[2]{Graduate University for advanced Studies (KEK), Ibaraki, Japan}
%
%
%
%
%
\begin{abstract}

A time projection chamber (TPC) is a strong candidate for the central tracker
of the international linear collider (ILC) experiment and we have been
conducting a series of cosmic ray experiments under a magnetic field up to 4 T,
using a small prototype TPC with a replaceable readout device:
multi-wire proportional chamber (MWPC) or gas electron multiplier (GEM).
We first confirmed that the MWPC readout could not be a fall-back option of
the ILC-TPC under a strong axial magnetic field of 4 T since its spatial
resolution suffered severely from the so called $E$ $\times$ $B$ effect
in the vicinity of the wire planes.
The GEM readout, on the other hand, was found to be virtually free from
the $E$ $\times$ $B$ effect as had been expected and gave the resolution
determined by the transverse diffusion of the drift electrons (diffusion limited).
Furthermore, GEMs allow a wider choice of gas mixtures than MWPCs.
Among the gases we tried so far a mixture of Ar-CF$_4$-isobutane, in which MWPCs
could be prone to discharges, seems promising as the operating gas of the ILC-TPC
because of its small diffusion constant especially under a strong magnetic field.
We report the measured drift properties of this mixture including the diffusion
constant as a function of the electric field and compare them with the
predictions of Magboltz.   
Also presented is the spatial resolution of a GEM-based ILC-TPC
estimated from the measurement with the prototype.

\end{abstract}
%
%
%
%
%
%

\end{frontmatter}


%
%
%
%
%
%

\section{Introduction}

One of the most important issues of the current high energy physics is to
find the Higgs boson and to reveal its nature. The LHC will most likely find a
Higgs candidate and the ILC is expected to follow it to complete the
mission~\cite{ILC1}\cite{ILC2}.
To study the Higgs properties in detail we need a high performance central tracker.

A TPC is a natural candidate for the ILC central tracker because of its very good
performance in the past collider experiments~\cite{Mike}.
At the ILC, however, we need the highest possible tracking efficiency in a jetty
environment and a momentum resolution one order-of-magnitude better
than those in the past.
This requires very high 3-dimensional granularity (a small voxel size) and
a space point resolution in the $r-\phi$ plane better than 100 $\mu$m throughout
the sensitive volume.  
In order to realize a TPC with such unprecedented performance, intense R\&D programs
are now on going in an international framework called the LC-TPC collaboration.
Three technologies have been considered as the gas amplification device for the
ILC-TPC: MWPC, MicroMEGAS~\cite{MicroMEGAS} and GEM~\cite{GEM}.

In order to compare the different readout technologies, we prepared a small
prototype TPC with a replaceable readout plane consisting of a gas amplification
device and a pad plane.
The results of the beam test at KEK for a MicroMEGAS readout using a gas mixture of
Ar-isobutane(5\%) have already been published~\cite{Paul}.
In this paper our results of cosmic ray tests for an MWPC readout using
a TDR gas (Ar-CH$_4$(5\%)-CO$_2$(2\%))~\cite{ILC1} and for a triple GEM readout
operated in a T2K gas (Ar-CF$_4$(3\%)-isobutane(2\%))~\cite{T2K} are presented.

\section{Experimental Setup}

\begin{figure}[hbt]
\begin{center}
\includegraphics*[scale=0.50]{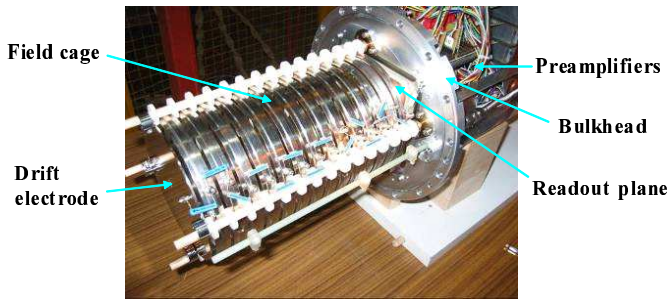}
\end{center}
\caption{\label{fig1}
Photograph of the prototype just before installation into the gas vessel.
}
\end{figure}

The GEM-based prototype is described in Ref.~\cite{Vienna}, where the preliminary
results of the beam test for the GEM readout using a TDR gas or
a P5 gas (Ar-CH$_4$(5\%)) are presented.
The sensitive volume of the prototype was $\sim$ 260 mm in length and 152 mm
in diameter.
The size of the pad plane was 100 mm $\times$ 100 mm.
The induction gap of the triple GEM stack was increased later from 1 mm to 1.5 mm
for the cosmic ray tests using a T2K gas, in order to increase the charge spread
after gas amplification, while the transfer gaps (1.5 mm) were maintained.
A picture of the prototype is shown in Fig.~\ref{fig1}.

For the MWPC readout the GEM foils were replaced by two wire planes~\cite{Ackermann}.
The sense-wire plane consisted of 20 $\mu$m thick wires spaced with 2 mm pitch
and was placed 1 mm above the readout pad plane, on which pad rows were arranged
in parallel to the wires.
The drift region was delimited by a grid wire plane.
The grid wire plane consisted of 127 $\mu$m thick wires strung 1 mm above the
sense-wire plane with 2 mm spacing.
It should be noted that the sense-wire pitch and the wires-to-pads distance were
kept small as compared to those in conventional TPCs in order to obtain the best
spatial resolution and the highest two-track resolving power achievable with a
practical MWPC readout\footnote
{
The width of the pad response function ($\sigma$) was about 1.3 mm.
}.
The pad pitch was increased to 2.3 mm,
from 1.27 mm in the case of the GEM readout\footnote
{
The neighboring pad rows were staggered by half a pad pitch in 
the GEM readout.
},
corresponding to the increase in the width of the pad response function
while the pad-row pitch (6.3 mm) was retained.

The MWPC-TPC was operated inside a 5-T solenoid at DESY and the GEM-TPC data was
taken inside a 1-T solenoid at the KEK cryogenic center. 
A pair of long scintillation counters was used for cosmic ray triggers.
The readout electronics system for the ALEPH TPC~\cite{ALEPH} was used for
the chamber signal processing and the traditional MultiFit (DoubleFit)
program~\cite{MultiFit} was used for the data analysis of the cosmic ray events.

The hit coordinate along a pad row was determined by a simple charge centroid
(barycenter) method using the pad signals within a charge cluster in the row.
Among the tracks recorded in the TPC, those which were nearly perpendicular to the
pad rows (sense wires) and parallel to the readout plane were selected
for later analysis.

\section{MWPC Readout}

\begin{figure}[hbt]
\begin{center}
\includegraphics*[scale=0.4]{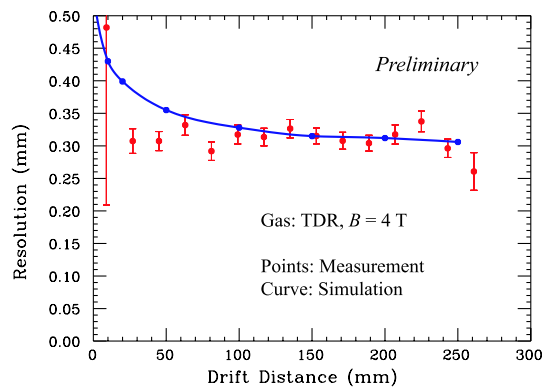}
\end{center}
\caption{\label{fig2}
Spatial resolution measured with the MWPC readout in a magnetic field of 4 T. 
The solid line shows the result of the simulation.
}
\end{figure}

The spatial resolution as a function of the drift distance ($z$) obtained under
a magnetic field ($B$) of 4 T is shown in Fig.~\ref{fig2}.
The sense-wire potential was set to 1200 V while the grid wires were grounded.
The drift field strength ($E_{\rm d}$) was 220 V/cm, at which the transverse
diffusion constant ($D$) of drift electrons is about 60 $\mu$m/$\sqrt{\rm cm}$.  
Also shown in the figure is the result of a Monte Carlo simulation.
The simulation took into account the primary ionization statistics of incident
tracks, the diffusion of drift electrons, the lateral displacement of the drift
electrons (charge spread) near the wire planes due to the $E \times B$ effect,
and finally the avalanche fluctuation in gas amplification.

\begin{figure}[hbt]
\begin{center}
\includegraphics*[scale=0.5]{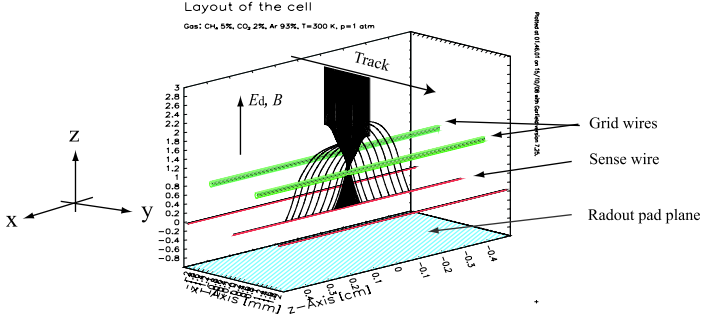}
\end{center}
\caption{\label{fig3}
Simulated drift electron paths without diffusion in TDR gas for a track
perpendicular to the wires ($B$ = 4 T).
}
\end{figure}

\begin{figure}[hbt]
\begin{center}
\includegraphics*[scale=0.45]{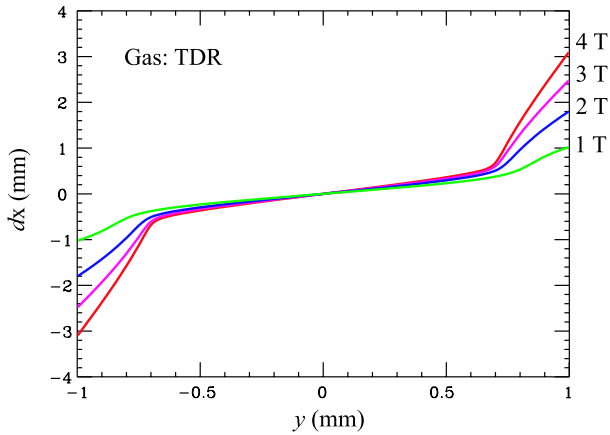}
\end{center}
\caption
{\label{fig4}
Displacement of the arrival position of drift electrons on the sense wire
as a function of the initial position along the track for $B$ = 1, 2, 3 and 4 T.
The coordinate $y$ is measured from the point on the track right above the
sense wire. 
The increase of displacement ($dx$), especially for large $y$, degrades the
spatial resolution significantly.
}
\end{figure}

The simulation program Garfield/Magboltz~\cite{Biagi} was used to estimate the
displacement of the drift electrons along the wire direction ($dx$) due to
the $E \times B$ effect as a function of the initial position along the track ($y$)
(see Fig.~\ref{fig3} and \ref{fig4}).
The simulated resolution clearly shows the large contribution of the $E \times B$
effect at short drift distances and its reduction with increasing $z$ due to
the declustering effect caused by the diffusion of the drift electrons~\cite{Blum}.

\begin{figure}[hbt]
\begin{center}
\includegraphics*[scale=0.5]{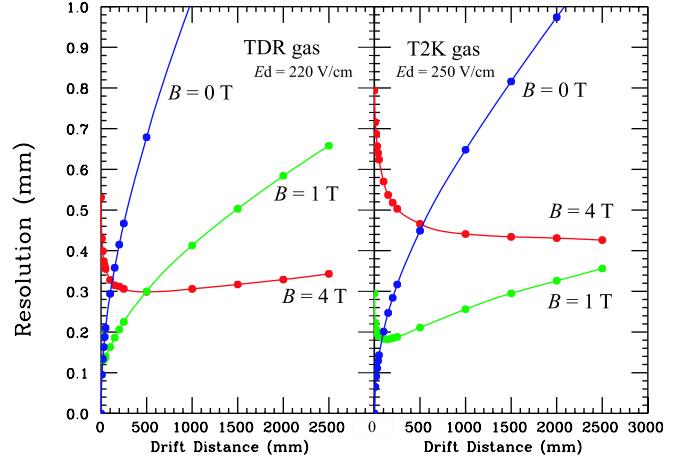}
\end{center}
\caption{\label{fig5}
Simulated spatial resolution of the ILC-TPC equipped with the MWPC readout
operated in TDR gas (left) and in T2K gas (right) for tracks perpendicular 
to the wires and the pad rows.
}
\end{figure}

Fig.~\ref{fig5} shows the simulated spatial resolution of the ILC-TPC
equipped with the MWPC readout for $B$ = 0, 1 and 4 T.
The resolution for $B$ = 4 T is dominated by the $E \times B$ effect throughout
the drift region in spite of its decrease at long drift distances
where the contribution of the electron diffusion increases.
The situation is worse for T2K gas because of the larger charge spread near the
wire planes caused by the $E \times B$ effect and
the smaller declustering effect due to its small diffusion constant
($D$ $\sim$ 30 $\mu$m/$\sqrt{\rm cm}$ at $E_{\rm d}$ = 250 V/cm and $B$ = 4 T).
Therefore T2K gas would not be a good choice for MWPC readout even if the MWPC
itself operated properly in this gas mixture. 

\section{GEM readout}

\begin{figure}[!h]
\begin{center}
\includegraphics*[scale=0.45]{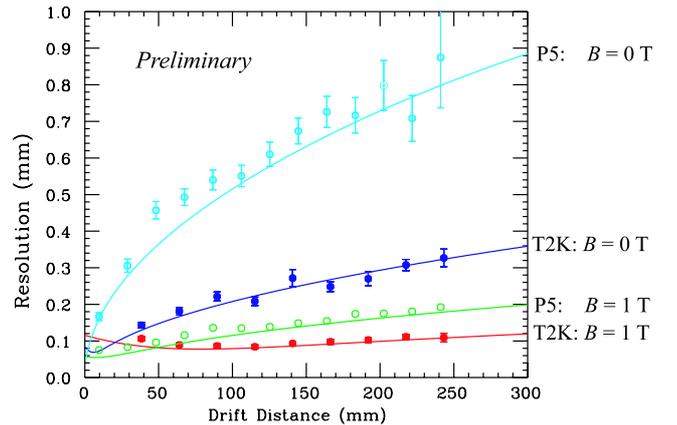}
\end{center}
\caption{\label{fig6}
Spatial resolution as a function of the drift distance obtained with a T2K gas 
($E_{\rm d}$ = 250 V/cm) for $B$ = 0 and 1 T. 
The resolutions obtained in the beam test with a P5 gas ($E_{\rm d}$ = 100 V/cm)
are also shown for comparison. 
}
\end{figure}

The spatial resolutions measured under magnetic fields of 0 and 1 T are shown
in Fig.~\ref{fig6}.
The high voltage applied across each GEM of the triple stack was 250 V and
the electric fields in the transfer and induction gaps were set to 1.6 kV/cm.
The solid curves in the figure are the results of analytic calculation~\cite{Paul}.
The resolution deterioration at short drift distances prominent for the T2K gas
at 1 T is not caused by an $E \times B$ effect but by the finite pitch of the
readout pads (1.27 mm), which is not small enough compared to the charge spread
after gas amplification in the GEM stack ($\sim$ 270 $\mu$m)~\cite{Paul}.

The spatial resolution ($\sigma_{\rm X}$) at long drift distances is given by
\begin{displaymath}
\sigma_{\rm X}^2 = \sigma_{\rm X0}^2 + D^2/N_{\rm eff} \cdot z\;,
\end{displaymath}
where $\sigma_{\rm X0}$ is the intrinsic resolution and $N_{\rm eff}$ is the
effective number of electrons, which is determined by the primary ionization
statistics and the avalanche fluctuation for a single drift electron~\cite{Neff}.

Since the diffusion constant is determined from the $z$-dependence of the
pad response width~\cite{Paul}\cite{Vienna} (or given by Magboltz)
one can evaluate $N_{\rm eff}$ from the $z$-dependence of the spatial resolution
at long drift distances.
The value of $N_{\rm eff}$ thus obtained was about 21, which is in good agreement
with a simple estimate of 22.2~\cite{Neff},
indicating that possible dissociative attachment of drift electrons by
CF$_4$ molecules~\cite{Christophorou1}\cite{Christophorou2}\cite{Anderson}
at the entrance to the GEM stack is not significant if any.

\begin{figure}[!h]
\begin{center}
\includegraphics*[scale=0.5]{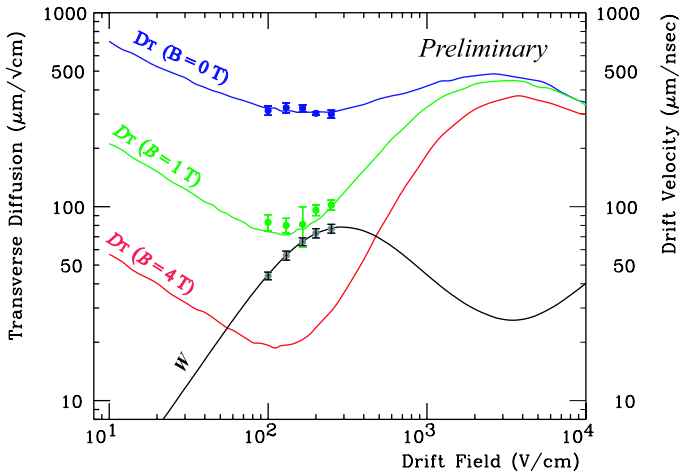}
\end{center}
\caption{\label{fig7}
Transverse diffusion constants ($D_{\rm T}$) and drift velocities ($W$) 
measured in T2K gas.
The curves represent the Magboltz predictions. .
}
\end{figure}

We also measured the diffusion constant and the drift velocity in a T2K gas as
functions of the drift field, covering from the diffusion minimum
($E_{\rm d}$ = 100 V/cm) to the drift velocity near-maximum ($E_{\rm d}$ = 250 V/cm).
The drift velocity was estimated from the full width of the drift time distribution
and the maximum drift length of the prototype.
The results are shown in Fig.~\ref{fig7} along with the corresponding Magboltz
predictions, which closely reproduce the data points. 
The reliability of Magboltz for this gas mixture was thus confirmed for the
practical range of the electric field for TPC operation.

\begin{figure}[!h]
\begin{center}
\includegraphics*[scale=0.45]{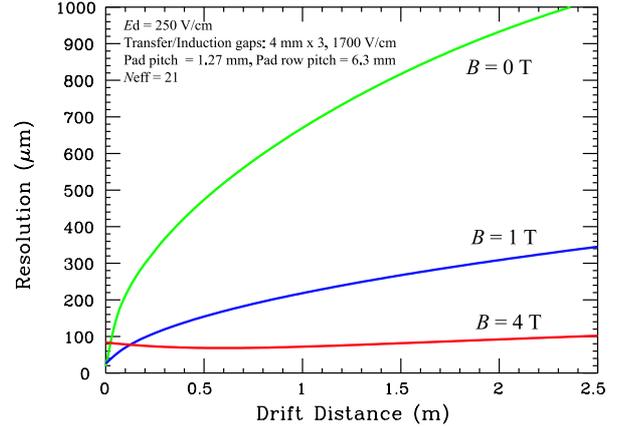}
\end{center}
\caption{\label{fig8}
Expected resolution of a GEM-based TPC operated in T2K gas
for tracks perpendicular to the pad rows.
}
\end{figure}

Fig.~\ref{fig8} shows the calculated resolution of a GEM-based ILC-TPC operated in a
T2K gas, assuming the diffusion constant given by Magboltz for $E_{\rm d}$ = 250 V/cm.

%

\section{Conclusion}

MWPC readout turned out to be inappropriate for a TPC operated in a strong
magnetic field because of a large $E \times B$ effect on its spatial resolution.
Therefore it can not be a fall-back option for the ILC-TPC. 
A GEM-based TPC using a T2K gas is a good candidate for the central tracker
of the ILC experiment.
It operates stably and is expected to give the required spatial resolution of
better than 100 $\mu$m throughout the sensitive volume under a 4 T magnetic field.

%

\section*{Acknowledgments}
We are grateful to the KEK cryogenic center and the IPNS cryogenic group
for the operation of the superconducting magnet. 
This study is supported in part by the Creative Scientific Research  
Grant no. 18GS0202 of the Japan Society for Promotion of Science.
%

\end{document}